\documentclass[lettersize,journal]{IEEEtran}
\usepackage{amsmath,amsfonts}
\usepackage{algorithm}
\usepackage{algpseudocode}
\usepackage{array}
\usepackage[caption=false,font=normalsize,labelfont=sf,textfont=sf]{subfig}
\usepackage{textcomp}
\usepackage{stfloats}
\usepackage{url}
\usepackage{verbatim}
\usepackage{graphicx}
\graphicspath{ {./Images/} }
\usepackage{cite}
\usepackage{xcolor}
\usepackage{booktabs}
\usepackage{pifont} 
\usepackage{tikz}
\usetikzlibrary{automata}
\usetikzlibrary{arrows.meta,positioning,fit}
\usepackage{tabularx}
\usepackage{mathtools}
\usepackage{siunitx}
\usepackage{makecell}

\begin{document}

\title{\emph{TLoRa}: Implementing TLS Over LoRa for Secure HTTP Communication in IoT}

\author{Atonu~Ghosh,~\IEEEmembership{Graduate~Student~Member,~IEEE,}
        Akhilesh~Mohanasundaram, Srishivanth~R~F,
        and~Sudip~Misra,~\IEEEmembership{Fellow~IEEE,~ACM}
\thanks{Atonu Ghosh, Akhilesh Mohanasundaram, Srishivanth R F, and Sudip Misra are with the Department of Computer Science and Engineering, Indian Institute of Technology Kharagpur, Kharagpur 721302, WB, India. (e-mail: atonughosh@outlook.com; akhileshmohan2005@gmail.com; srishivanth04@gmail.com; sudip\_misra@yahoo.com)}%
}

\markboth{Work in Progress}%
{Ghosh \MakeLowercase{\textit{et al.}}: TLoRa: Enabling Secure HTTP Communication Over LoRa Using TLS for Last-Mile IoT}


\maketitle

\begin{abstract}
We present \emph{TLoRa}, an end-to-end architecture for HTTPS communication over LoRa by integrating TCP tunneling and a complete TLS 1.3 handshake. It enables a seamless and secure communication channel between WiFi-enabled end devices and the Internet over LoRa using an End Hub (EH) and a Net Relay (NR). The EH tethers a WiFi hotspot and a captive portal for user devices to connect and request URLs. The EH forwards the requested URLs to the NR using a secure tunnel over LoRa. The NR, which acts as a server-side proxy, receives and resolves the request from the Internet-based server. It then relays back the encrypted response from the server over the same secure tunnel. \emph{TLoRa} operates in three phases -session setup, secure tunneling, and rendering. In the first phase, it manages the TCP socket and initiates the TLS handshake. In the second, it creates a secure tunnel and transfers encrypted TLS data over LoRa. Finally, it delivers the URL content to the user. \emph{TLoRa} also implements a lightweight TLS record reassembly layer and a queuing mechanism for session multiplexing. We evaluate \emph{TLoRa} on real hardware using multiple accesses to a web API. Results indicate that it provides a practical solution by successfully establishing a TLS session over LoRa in 9.9 seconds and takes 3.58 seconds to fulfill API requests. To the best of our knowledge, this is the first work to comprehensively design, implement, and evaluate the performance of HTTPS access over LoRa using full TLS.
\end{abstract}

\begin{IEEEkeywords}
TLS Over LoRa, API Over LoRa, HTTPS Over LoRa, Secure LoRa, Internet of Things, Web of Things.
\end{IEEEkeywords}

\section{Introduction}
\IEEEPARstart{L}{oRa} has proved its effectiveness in modern Internet of Things (IoT) applications by virtue of its long range and low power capabilities \cite{10457052, 10978509}. At the same time, IoT is evolving and gradually converging towards the Web of Things (WoT) paradigm. These operate beyond the traditional objective of just "pushing" the data to the destination and utilize the standard web protocols such as HTTP and REST \cite{9531380}. But the limited bandwidth of LoRa restricts the wider adoption and becomes a bottleneck in such request-response-based scenarios. LoRa becomes restrictive, especially in applications that require advanced security measures \cite{9993728}. This is primarily due to the need for substantial and systematic data exchange while implementing strong security measures such as certificate-based authentication. The exchange of digital certificates and the timed multi-step handshakes over LoRa require more efficient mechanisms. Numerous research efforts have been made in the recent past that aim to make LoRa communication more secure. However, they fail to enable secure access to web APIs, a critical aspect of modern IoT driven by complex web applications. Thus, creating a pressing need for solutions that can bridge the gap between the modern needs of IoT systems and the highly restrictive LoRa physical layer. 

LoRaWAN's dual-layer AES-128-based encryption mechanism mitigates the security issues to some extent, but these measures still fail to enable end-to-end secure web API access, an essential requirement of modern IoT \cite{10.1145/3543856}. This is due to the LoRaWAN's physical limitations and rigid architecture \cite{NOURA2020100303} which does not enable the end-nodes to communicate with the secure web APIs. Furthermore, it also deters the implementation of custom IoT web services. Alternative communication technologies such as NB-IoT, which support an IP stack, can enable secure API access. But it necessarily requires cellular infrastructure, a subscription plan, and consumes significantly more energy than LoRa \cite{10.1145/3536424, 9435286, 9514451}. Thus, making it inflexible and limiting its viability in remote and disconnected locations that lack the infrastructure. On the other hand, Sigfox not only provides ultra-low data rates but also requires a subscription from a service provider \cite{10084406, 9148645}.

\begin{figure}[h]
    \centering
    \includegraphics[width=0.85\columnwidth]{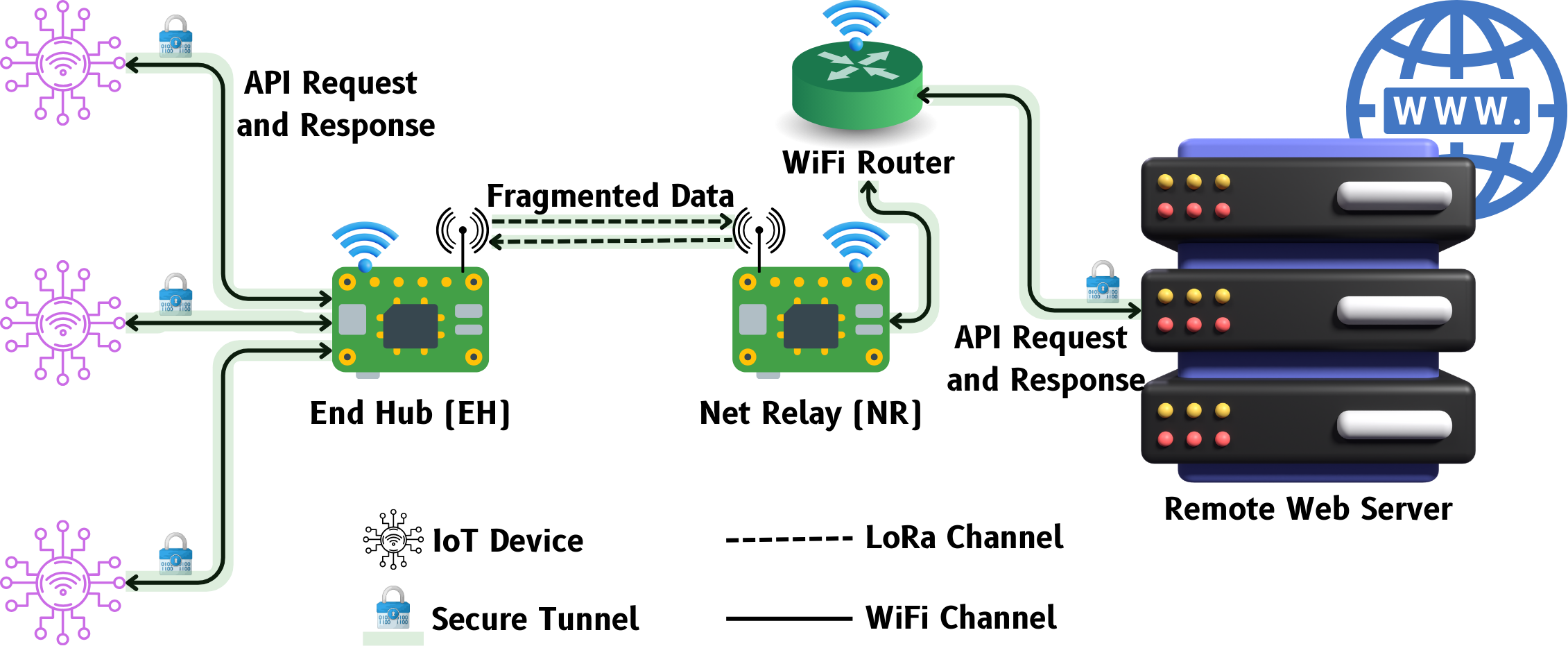}
    \caption{An end-to-end overview of the proposed architecture showing different components of the system and their interconnections.}
    \label{fig:overview}
\end{figure}

In this work, we propose \emph{TLoRa}, an end-to-end architecture that enables web API access directly from IoT nodes using the standard HTTPS protocol over the LoRa channel. It establishes a full TLS 1.3 handshake and creates a secure tunnel before allowing secure API access. Figure \ref{fig:overview} depicts a basic outline of the setup in an example topology of \emph{TLoRa}. The EH receives the requests from end devices over WiFi, processes them, and sends the requests to the NR over a LoRa channel. Once the NR receives the requests over LoRa, it processes and sends the requests to the web server over WiFi or Ethernet. The responses follow the reverse path and reach the end devices. As the network is inherently multi-hop and consists of different communication channels, \emph{TLoRa} employs appropriate methods to manage data exchange in such an environment. We discuss the detailed architecture and methods of \emph{TLoRa} in Section \ref{system_model}.

\subsection{Motivation}
With the proliferation of IoT in all aspects of human life, IoT systems now transport a growing volume of sensitive data. In addition, the requirement for ubiquity in IoT is unprecedented, and as a result, modern solutions are primarily Internet-oriented. Consequently, modern IoT deployments often demand secure, trustworthy, and energy-efficient enabling technologies. They must also provide greater flexibility in the application of sensitive business logic to data, while allowing control over both the data and the communication channel. The existing  LoRa-based solutions do not fully support these requirements and they do not enable secure and direct API access. Existing solutions, such as LoRaWAN and NB-IoT, are heavily dependent on the infrastructure of service providers and lack flexibility, hindering future innovations. Some major lacunae of the existing systems, which motivated us to pursue this work of enabling secure web API access over LoRa using TLS, are:
\begin{enumerate}
    \item Present-day solutions suffer due to the physical layer limitations and architectural rigidity. They do not allow LoRa devices to securely access/interact with web APIs over TLS/HTTPS, which are session-based and synchronous protocols. This prevents the integration of the constrained IoT environment with modern applications.
    
    \item The state-of-the-art fails to provide a framework that optimizes the data-intensive, timed, and multi-step handshakes of modern security protocols such as TLS over a bandwidth-constrained LoRa physical layer. Rademacher et al. \cite{9861875} quantified the limitations of implementing TLS over LoRaWAN through their airtime model. They found that a full TLS handshake generates $3-6$ kB of data and it becomes an overhead when combined with the strict duty cycle requirements of LoRaWAN. Their conclusion that downlink communication becomes the bottleneck necessitates a new approach that can make such session-oriented security protocols possible on LoRa.
    
    \item The existing solution approaches do not address the issue of end-to-end secure and direct API access over LoRa using a full TLS handshake \cite{10.1145/3747295, 10.1145/3530190.3534807, ghosh2025loraconnect}.
\end{enumerate}

\subsection{Contribution}
Towards the seamless integration of LoRa-based IoT environments with the modern and complex web applications, we propose an IoT architecture for direct access and interaction with web APIs. The proposed architecture overcomes the limitations of the state-of-the-art and does not require any modification to end-devices. The contributions of our work are:

\begin{enumerate}
    \item We propose a highly flexible and extensible proxy-based architecture to achieve a complete TLS handshake via the TCP tunnel over the LoRa backhaul. It does not necessitate any modification to the user's application or end devices. 

    \item We introduce a Finite State Machine (FSM)-based method for predictable state transitions. This improves the system's reliability. The FSM further allows strict and accurate control of the sequential events in the system.

    \item We design and implement a packet manipulation mechanism across layers with a custom data transport protocol above the data slicing mechanism in \cite{ghosh2025loraconnect}. Our method performs TCP header manipulation to correct TCP timestamps in the SYN-ACK packet. This keeps the session integrity intact over the LoRa channel.

    \item We implement and validate the proposed system on actual lab-scale hardware prototype. We also evaluate it based on various key performance indicators (KPIs) such as latency, packet delivery ratio (PDR), and power consumption. Results indicate a successful and full TLS handshake over LoRa.
\end{enumerate}

\section{Related Work}
The severely constrained nature of LPWAN, like LoRa, presents a severe challenge in its integration with the public Internet. Protocols like HTTPS require low-latency and high-bandwidth connections, which is in stark contrast with the low data rate and high latency nature of LPWANs. As a result, the existing research has largely focused on the adaptation layer, compression schemes, and alternative lightweight protocols, which we discuss in detail.

\subsection{Security in LoRaWAN}
Despite the latest release of security specifications that implement a layered AES-128-bit encryption mechanism, LoRaWAN still suffers heavily from a lack of advanced security measures \cite{10.1145/3561973}. Abboud et al. \cite{10376065} enhanced this method using AES-256. They proved the efficiency of their approach in their work, which consumed marginal resources but ensured greater security. They observed that the cipher-text transmission only increased to $4.1$  ms from $2.66$ ms as with AES-128. Hayati et al. \cite{9709341} addressed a more fundamental security issue in LoRaWAN where the root keys were static. They proposed a method to keep the root key constantly updated using the CTR AES DRBG 128 algorithm which secured the previous and future keys even if a root key was compromised. Chang et al. \cite{chang2025low} experimented with a hybrid model for the key exchange process and introduced an RSA-AES algorithm specifically tuned for resource-constrained IoT devices. In this, they modified the standard algorithms using a triple-prime system for RSA and reduced the AES operation to have only seven rounds. A multi-threaded design was used where the stronger and slower asymmetric M-RSA was used to encrypt and transmit the symmetric S-AES session key. The S-AES was then used to encrypt the payload. In a similar work, Akram et al. \cite{11104491} proposed a method for protecting the privacy in the authentication process of LoRaWAN. They used Physically Unclonable Functions (PUFs) instead of storing the static keys on the device. This method provided higher resilience and prevented key extraction and device cloning. Recently, Vikash et al. \cite{10934001} proposed a Bit-Mapping-based security-aware MAC protocol that distinguished between critical and non-critical nodes in a deployment. The protocol dynamically adjusted the security levels, i.e., it automatically applied the standard AES-128 for non-critical data and applied the stronger AES-256 for critical ones. They further implemented channel prioritization and energy optimization to maintain the energy efficiency of the network. 

\newcommand{\cmark}{\textcolor{green!70!black}{\ding{51}}}
\newcommand{\xmark}{\textcolor{red}{\ding{55}}}

\begin{table*}[t] 
  \centering
    \caption{Comparison of our work on implementing TLS over LoRa with related work}
    \label{related_work}
    \scriptsize
  \begin{tabular}{c c c c c c c c c}
    \toprule
    \thead{Works} & \thead{IP Communication} & \thead{Lightweight / Optimized Crypto} & \thead{Multi-hop Support} & \thead{Hardware-based} & \thead{Full TLS \& HTTPS Over LoRa}\\
    \midrule
    Paris et al.\: \cite{paris2023implementation} & \xmark & \xmark & \xmark & \cmark & \xmark\\
    Bhardwaj et al.\: \cite{10911060} & \xmark & \xmark & \xmark & \cmark & \xmark\\
    Cilfone et al.\: \cite{10415221} & \cmark & \xmark & \xmark & \cmark & \xmark\\
    Aramaki et al.\: \cite{10621878} & \cmark & \xmark & \cmark & \cmark & \xmark\\
    Abboud et al.\: \cite{10376065} & \xmark & \cmark & \xmark & \xmark & \xmark\\
    Hayati et al.\: \cite{9709341} & \xmark & \cmark & \xmark & \cmark & \xmark\\
    \textbf{\textit{TLoRa (Our Work)}} & \textbf{\cmark} & \textbf{\cmark} & \textbf{\cmark} & \textbf{\cmark} & \textbf{\cmark}\\
    \bottomrule
  \end{tabular}
\end{table*}

\subsection{Lightweight Security Mechanisms for IoT}
The implementation of full TLS, an essential component of HTTPS, on the heavily resource-constrained LoRa-based IoT systems, was widely recognized as infeasible \cite{10.1007/978-3-030-61078-4_14, 9508382}. As a result, researchers came up with lightweight solutions. Paris et al. \cite{paris2023implementation} implemented TLS/SSL with MQTT on ESP32 and Raspberry Pi devices. They observed the energy consumption to increase four times and concluded that full TLS implementation is impractical for most battery-powered IoT nodes. Furthermore, they observed high overhead and longer execution times. In spite of the challenges, Bhardwaj et al. \cite{10911060} implemented TLS to secure application layer protocols like MQTT in LoRa-based military and industrial systems. Due to these limitations, researchers developed adaptations and optimizations of the TLS stack. Datagram Transport Layer Security (DTLS) is one of the major adaptations for IoT. It is well-suited for connectionless setups that primarily rely on User Datagram Protocol (UDP). Researchers have also put efforts into making TLS efficient for resource-constrained devices because even DTLS can be too resource-intensive for constrained IoT systems \cite{8962334}. Bodenhausen et al. \cite{bodenhausen2025bidirectionaltlshandshakecaching} proposed a method for caching TLS certificates for both the client and the server. They achieved better performance as it only required a small fingerprint of the cached certificate to be exchanged in the subsequent handshakes. They observed reductions by 61.1\% and  8.5\% in bandwidth consumption and computational overhead, respectively. Furthermore, for efficient computation of Elliptic Curve Cryptography (ECC), Mao et al. \cite{10189076} proposed a RISC-V-based lightweight system and used a hardware accelerator. They observed significant improvements and energy efficiency and concluded that the proposed system is applicable in practical IoT deployments.

\subsection{IP-based Communication in LPWANs}
LoRaWAN's non-IP-based architecture primarily prevents the application of standard security protocols such as TLS. Several research efforts have gone into building adaptation layers that make IP-based communication over LoRa possible. Cilfone et al. \cite{10415221} proposed a framework using a container-based virtualization to connect non-IP-based LoRaWAN end devices to an IP-based network using CoAP. The system created digital twins of each node and did not require any modification to the underlying LoRaWAN stack. The researchers further demonstrated the proposed framework using laptops and Raspberry Pi platforms. Other research efforts have also gone into this, where the network layer was extended directly into the LoRa devices. Inspired by 6LoWPAN, Herrero \cite{HERRERO2023100647} proposed a mechanism to enable IPv6 in LoRa topologies. They chunked and compressed the header to shrink the IP datagrams. The researcher validated the approach through experiments using Raspberry Pi and AWS cloud. Additionally, towards this end, Aramaki et al. \cite{10621878} experimented to evaluate multi-hop TCP/IP communication over LoRa using IP2LoRa. The nodes in their experiment consisted of two LoRa transceivers for simultaneous transmission and reception. They concluded that the UDP throughput remained consistent over hops, but it dropped to about half for TCP. In spite of the drop in throughput, they found the TCP/IP communication to be stable. As routing is an essential component of multi-hop networks, Ghosh et al. \cite{10142009} proposed a routing scheme for LoRa backhaul networks for TCP/IP communication. They also evaluated their system and the proposed routing mechanism on actual hardware, confirming applicability. Ghosh et al. \cite{ghosh2025loraconnectunlockinghttppotential} proposed a system for enabling HTTP access directly on LoRa. They also proposed a message fragmentation and reassembly mechanism. They validated their proposed system's applicability and feasibility through experimental results.

\subsection{Synthesis}
These research efforts together highlight the varied research efforts that have gone towards making LPWANs more secure and reliable. Significant research efforts have been made to strengthen the core cryptographic methods. The state-of-the-art highlights the architectural bottleneck of LoRaWAN becoming the major obstacle in the implementation of an end-to-end Internet-standard security. Despite the challenges, recent works have proposed IP adaptation layers to enable IP communication on LPWANs. But there is a lack of practical end-to-end implementation of TLS directly on LoRa infrastructure that enables direct API access using HTTPS. To address this lacuna, we present \emph{TLoRa}. It is the first end-to-end design, implementation, and performance evaluation of the standard HTTPS protocol using full TLS over LoRa.

\section{System Model} \label{system_model}

\subsection{System Architecture}
The proposed \emph{TLoRa} system, as depicted in Fig. \ref{fig:overview} comprises -

\emph{End Devices (ED)}: These are the devices in an IoT deployment that initiate HTTPS requests. Typically, the end devices are resource-constrained devices such as sensors and actuators. But the \emph{TLoRa} architecture does not mandate their resource-constrained nature.  These can also be powerful devices like laptops or smartphones. The only assumption made in the proposed system is that these devices have the necessary protocol stack to generate an HTTPS request over WiFi. These devices are oblivious to the remainder of the system and the LoRa backhaul network.

\begin{figure}[h]
    \centering
    \includegraphics[width=0.8\columnwidth]{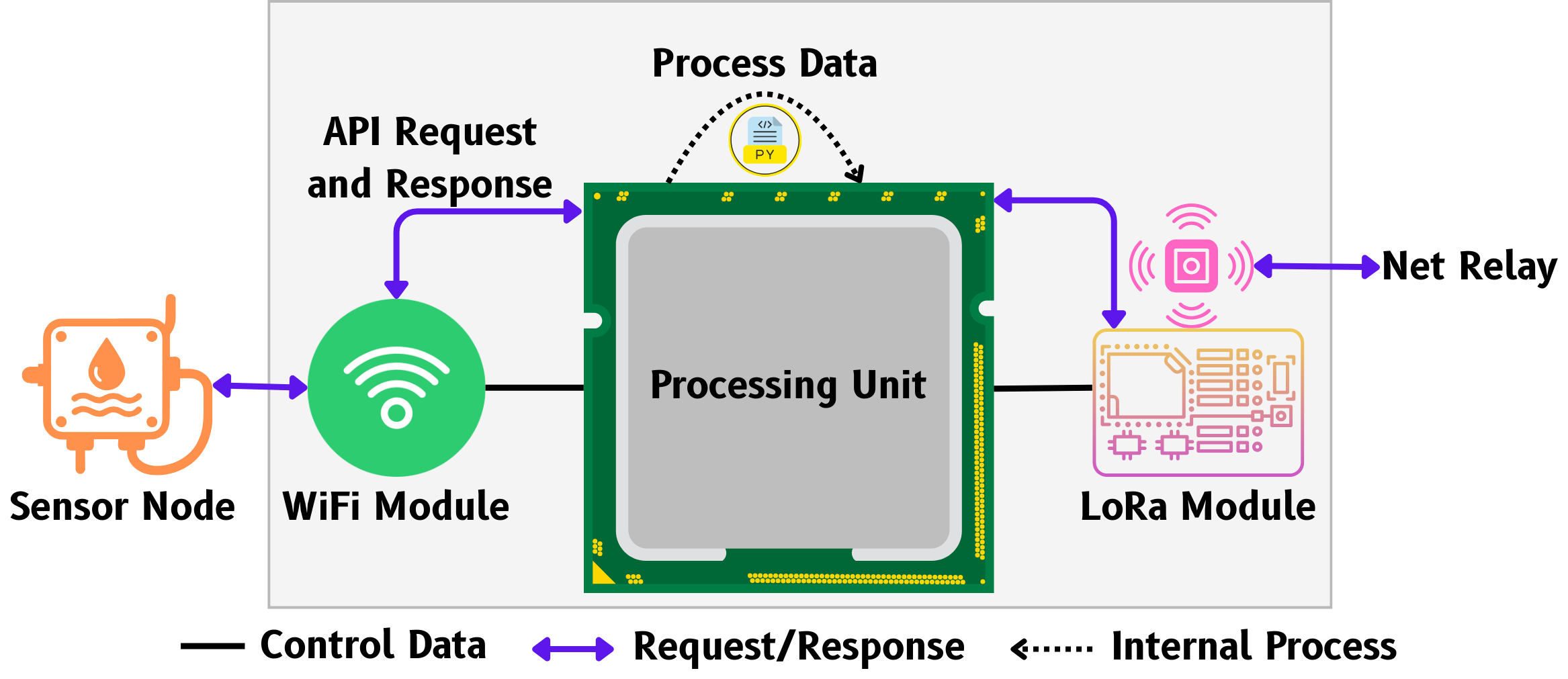}
    \caption{Internal architecture of End Hub (EH) and Net Relay (NR) in the proposed \emph{TLoRa} system.}
    \label{fig:endhub}
\end{figure}

\emph{End Hub (EH)}: Figure \ref{fig:endhub} depicts the internal architecture of the EH. It creates a bridge between the ED and the LoRa communication channel for seamless data flow.  It utilizes the WiFi module and software routines to create a WiFi hotspot-based Wireless Local Area Network (WLAN) with no Internet connectivity to which the EDs connect.

\emph{Net Relay (NR)}: The internal architecture of the NR is exactly the same as the EH. But it receives the requests from the EH on the LoRa channel, processes them, and sends the requests to the web server over WiFi. The WiFi in this case has Internet connectivity.

\emph{Web Server}: It is a standard web server accessible over the public Internet. \emph{TLoRa} only requires an API exposed using TLS and does not necessitate any modification in the server's application software.

\subsection{Network Architecture}
\emph{TLoRa} presents a flexible and extensible network design with only one requirement that the EH and NR are communicating. The end nodes can form any topology with the EH with one or more hops. They may form a star, a star of stars, a ring, or a bus topology before reaching the EH. In one of its forms as depicted in Fig. \ref{fig:overview}, \emph{TLoRa} consists of $n$ distributed sensor nodes $S = {s_1, s_2, \dots , s_n}$, one EH $\Psi_c$, one NR $\Psi_s$, and a web server $\Phi$. The bidirectional \emph{TLoRa} network ($\Pi$) is represented as an undirected graph  $\Pi = (\Omega, \Upsilon)$ where, $\Omega$ is the set of vertices (devices in the network) (Eq. \ref{vertices}) and $\Upsilon$ is the set of edges (communication links) (Eq. \ref{links}).

\begin{equation}
\label{vertices}
    \Omega = S \cup \{\Psi_c, \Psi_s, \Phi\}
\end{equation}

\begin{equation}
\label{links}
    \Upsilon = \Upsilon_\text{sensor} \cup \Upsilon_\text{backbone}
\end{equation}

where $\Upsilon_{\text{sensor}}$ are the communication links in the sensor network topology which is flexible in nature. $\Upsilon_{\text{backbone}}$ are the communication links between $\Psi_c$ and $\Psi_s$ and between $\Psi_s$ and $\Phi$ in the \emph{TLoRa} network. We further represent these segments in their constituent elements as -

\begin{equation}
    \label{flexibility}
    \Upsilon_{\text{sensor}} \subseteq \{\{u, v\} \mid u, v \in S, u \neq v\} \cup \{\{s, \Psi_c\} \mid s \in S\}
\end{equation}

\begin{equation}
\label{network_breakdown}
    \Upsilon_{\text{backbone}} = \{\{\Psi_c, \Psi_s\}, \{\Psi_s, \Phi\}\}  
\end{equation}

The only constraint of each sensor node sending data to $\Psi_c$ is represented in the \emph{TLoRa} system as -

\begin{equation}
\label{network_constraint}
    \forall s_i \in S,  \exists \text{ a path } (s_i, v_1, \dots, v_k, \Psi_c) \text{ in } \Pi
\end{equation}

We finally represent the end-to-end data flow path ($\Theta_{s_i}$) from any sensor $s_i$ to the server as  $\Theta_{s_i} : s_i \leadsto \Psi_c \rightarrow \Psi_s \rightarrow \Phi$ where $u \leadsto v$ denotes a single or multi-hop path between any two arbitrary end devices $u$ and $v$, and $u \rightarrow v$ denotes a direct link in the network backbone.

\subsection{Communication Model}
The proposed \emph{TLoRa} architecture is highly flexible and supports event-driven, time-driven, and query-driven communication in the network. This work evaluates the event-driven communication model, as the Key Performance Indicators (KPIs) for all the communication models are very similar. The end-to-end data flow in \emph{TLoRa} along the path $\Theta_{s_i}$ uses heterogeneous communication stacks. The sensor nodes ($S$) and the EH ($\Psi_c$) communicate over WiFi where each sensor node ($s_i$) sends data ($D(s_i)$) using TCP/IP. In this part of the network, the flexibility is defined as in Eq. \ref{flexibility}. The standard WiFi routing and Medium Access Control mechanisms regulate the data flow. 

The backbone LoRa link ($\Psi_c \leftrightarrow \Psi_s$) implements a secure tunnel over its constrained bandwidth. To efficiently manage the large messages \emph{TLoRa} implementats a message slicing and reassembly mechanism in both $\Psi_c$ and $\Psi_s$. The segment size in our implementation os set to $200$ bytes ($L_{max})$. For a message of size $|D(s_i)|$, the number of segments ($k_i$) can be obtained as $k_i = \left\lceil \frac{|D(s_i)|}{L_{\text{max}}} \right\rceil$.

Equation \ref{split} represents the fragmentation function $\mathcal{F}$ mapping the payload to the ordered sequence of chunks $(c_{i,1}, c_{i,2}, \ldots, c_{i,k_i})$.

\begin{equation}
\label{split}
\mathcal{F} : D(s_i) \;\mapsto\; (c_{i,1}, c_{i,2}, \ldots, c_{i,k_i})
\end{equation}

where each chunk ($c_{i,j}$ for $j \in \{1, \dots, k_i -1 \}$) is of size $|c_{i,j}| = L_{max}$ and the last chunk is of size $|c_{i,j}| = |D(s_i)| (\text{mod} L_{max})$ when modulus is $\neq 0$. The original payload can be represented by concatenation ($\oplus$) of the chunks in Eq. \ref{split} as $D(s_i) = c_{i,1} \oplus c_{i,2} \oplus \dots \oplus c_{i,k_i}$

Before a chunk ($c_{i,j}$) is transmitted, \emph{TLoRa} wraps it into packet ($P_{X,j}$). Meta data such as Payload ID ($\lambda_i$), Total Chunks ($k_i$), and Chunk Index ($j$) are added to the header. \emph{TLoRa} also implements an acknowledgment (ack) and retransmission mechanism where a packet gets resent if an ack is not received within a specific period. The user can set the number of retries and the timeout duration as per the Quality-of-Service (QoS) requirements. The receiver (either $\Psi_c$ or $\Psi_s$) reassembles the fragmented packets to get back the original message. In the reassembly process it uses the metadata in the packet header. This reassembly can be represented as a function $\mathcal{R}$, the inverse of fragmentation as in Eq. \ref{split}. The function $\mathcal{R}$ is represented as -

\begin{equation}
\label{reassemble}
\hat{D}(s_i) = \mathcal{R}(c_{i,1}, c_{i,2}, \ldots, c_{i,k_i})
= \bigoplus_{j=1}^{k_i} c_{i,j}
\end{equation}

where $\hat{D}(s_i)$ is the reassembled payload and $\hat{D}(s_i) = D(s_i)$ in a successful transmission. In a bidirectional communication, the roles of the slicer and reassembler are frequently reversed.

\begin{figure*}
    \centering
    \includegraphics[width=0.9\textwidth]{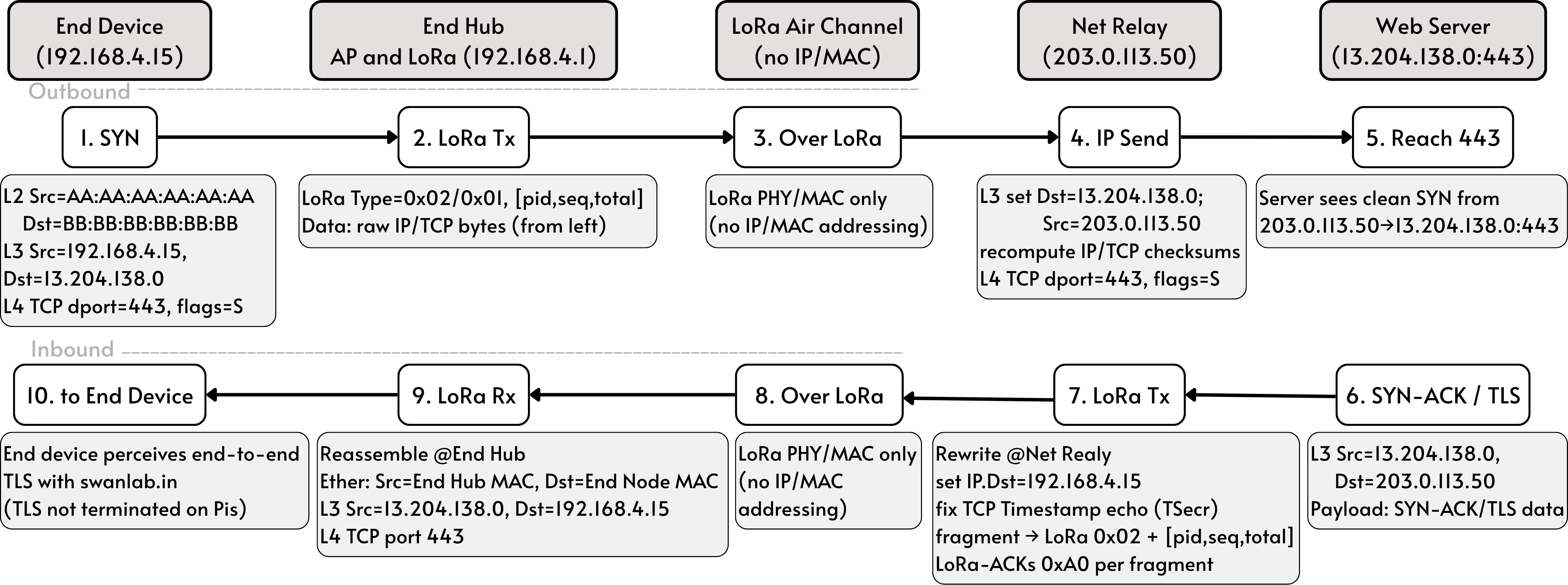}
    \caption{End-to-end packet journey with per-hop header changes and LoRa framing in \emph{TLoRa}.}
    \label{fig:packettransformation}
\end{figure*}

A WiFi-based Internet connection is used in the final segment ($\Psi_s \rightarrow \Phi$) of the \emph{TLoRa} network. But this segment does not create a separate web request. Instead, it completes the transparent TCP tunnel.  Figure \ref{fig:packettransformation} shows how a packet gets transformed at multiple stages in \emph{TLoRa}. EH ($\Psi_c$) treats a full IP packet from an end device as the payload ($D(s_i)$ which is chunked for transmission over LoRa. The NR ($\Psi_s$) reassembles these chunks and performs Network Address Translation (NAT) and replaces the end device's private IP with its own public IP before forwarding the packet to the web server. On the other hand, for a response packet, the destination IP is rewritten to the original end device's private IP. A key transformation in this process is the TCP timestamp correction. The TCP options of the SYN-ACK packets are modified by the NR to set the end device's original timestamp value. This is extremely critical to ensure that the TCP handshake is compatible with modern servers. This correction makes the whole process seamless and transparent. This makes the end-to-end proxy mechanism transparent in \emph{TLoRa}.

\subsection{TLoRa State Machine}
TLS handshake requires the execution of the stages in precise sequential steps. The LoRa channel, being very restrictive and susceptible to collisions and data losses, requires disciplined management. Hence, \emph{TLoRa} implements a Finite State Machine model (FSM) for both $\Psi_c$ and $\Psi_s$. This enables an organized flow of system control, efficient management of the half-duplex channel, smooth recovery in case of errors, enhanced scalability, and determinism.

\subsubsection{End Hub State Machine ($\mathcal{M}_{\Psi_c}$)}
We model $\mathcal{M}_{\Psi_c}$ FSM as a 5-tuple such that $\mathcal{M}_{\Psi_c} = (Q_c, \Sigma_c, \delta_c, q_{0_c}, F_c )$ \cite{10580939}, with -

\vspace{1mm}

$Q_c = \{C_0, C_1, C_2, C_3, C_4\}$

$\Sigma_c = e_0,e_1,e_2,e_3,e_4,e_5,e_6, e_7, e_8$

$q_{0_c} = C_0$

$F_c = C_0$

$\delta_c$ is the transition function detailed in Table~\ref{tab:client_delta}.

\begin{figure}[H]
    \centering
    \includegraphics[width=0.75\columnwidth]{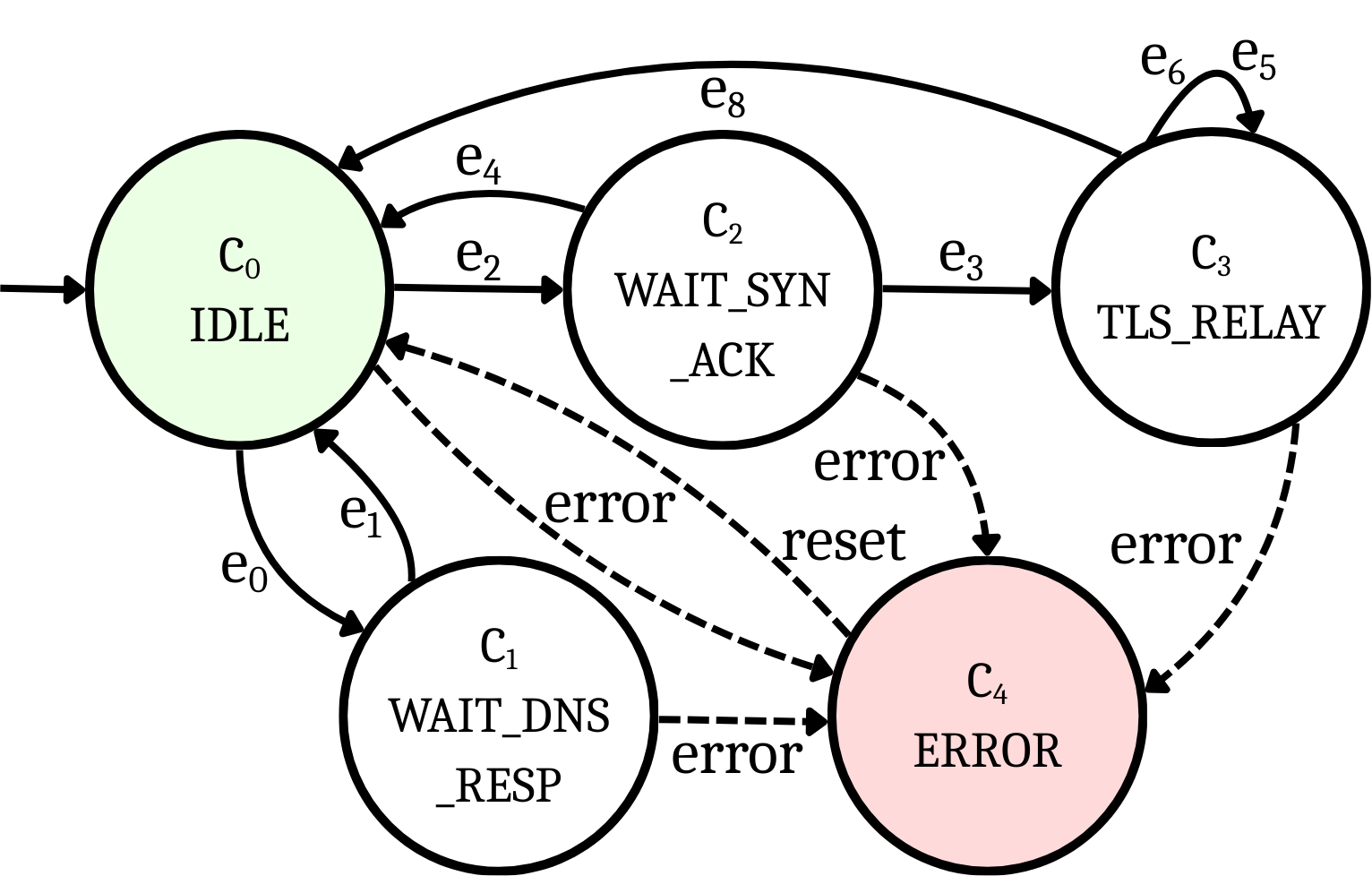}
    \caption{End Hub State Machine}
    \label{fig:clientfsm}
\end{figure}

\vspace{1mm}

where, $C_0$ : \emph{IDLE} (sniffing new DNS query or new connection), $C_1$ : \emph{WAIT\_DNS\_RESP} (waiting for LoRa resolved IP over LoRa), $C_2$ : \emph{WAIT\_SYN\_ACK} (waiting for LoRa response), $C_3$ : \emph{TLS\_RELAY} (Handshake complete, forwarding data), $C_4$: \emph{ERROR}.

$e_0$: Sniffed new DNS query, $e_1$: LoRa message with resolved IP, $e_2$: Sniffed new local TCP SYN, $e_3$: LoRa message with SYN-ACK received, $e_4$: Timeout waiting for LoRa SYN-ACK, $e_5$: Sniffed outgoing local TLS packet, $e_6$: LoRa message with TLS fragment received, $e_7$: LoRa ACK for a sent fragment received, and $e_8$: Session timeout, error, or FIN received. Figure \ref{fig:clientfsm} depicts the FSM and Table~\ref{tab:client_delta} depicts the state transitions.

\begin{table}[h!]
        \caption{Transition Function ($\delta_C$) for the Client FSM.}
        \label{tab:client_delta}
        \scriptsize
        \begin{tabularx}{\columnwidth}{l l l X}
            \toprule
            \textbf{State} & \textbf{Event} & \textbf{Next State} & \textbf{Action(s)} \\
            \midrule
            $C_0$ & $e_0$ & $C_1$ & Send domain name over LoRa \\
            $C_1$ & $e_1$ & $C_0$ & Spoof the DNS response to client \\
            $C_0$ & $e_2$ & $C_2$ & Send SYN over LoRa \\
            $C_2$ & $e_3$ & $C_3$ & Send final ACK over LoRa \\
            $C_2$ & $e_4$ & $C_0$ & Send FIN-ACK to client, log error \\
            $C_3$ & $e_5$ & $C_3$ & Chunk \& send over LoRa with ACK \\
            $C_3$ & $e_6$ & $C_3$ & Reassemble and forward to client \\
            $C_3$ & $e_6$ & $C_4$ & Send FIN-ACK to client \& cleanup \\
            $Any$ & - & $C_4$ & Cleanup and reset to IDLE \\
            \bottomrule
        \end{tabularx}
\end{table}

\subsubsection{Net Relay State Machine ($\mathcal{M}_{\Psi_s}$)}
We model $\mathcal{M}_{\Psi_s}$ FSM as a 5-tuple such that $\mathcal{M}_{\Psi_s} = (Q_s, \Sigma_s, \delta_s, q_{0_s}, F_s )$ \cite{10580939}, with -

\vspace{1mm}

$Q_s = \{S_0, S_1, S_2, S_3, S_4\}$

$\Sigma_s = e_0,e_1,e_2,e_3,e_4,e_5,e_6, e_7$

$q_{0_s} = S_0$

$F_s = S_0$

$\delta_s$ is the transition function detailed in Table~\ref{tab:client_delta}.

\vspace{1mm}

where, $S_0$ : \emph{IDLE} / \emph{WAIT\_DNS\_QRY},  $S_1$ : \emph{WAIT\_SYN}, $S_2$ : \emph{WAIT\_ACK}, $S_3$ : \emph{TLS\_RELAY}, $S_4$: \emph{ERROR}.

\begin{figure}[H]
    \centering
    \includegraphics[width=0.8\columnwidth]{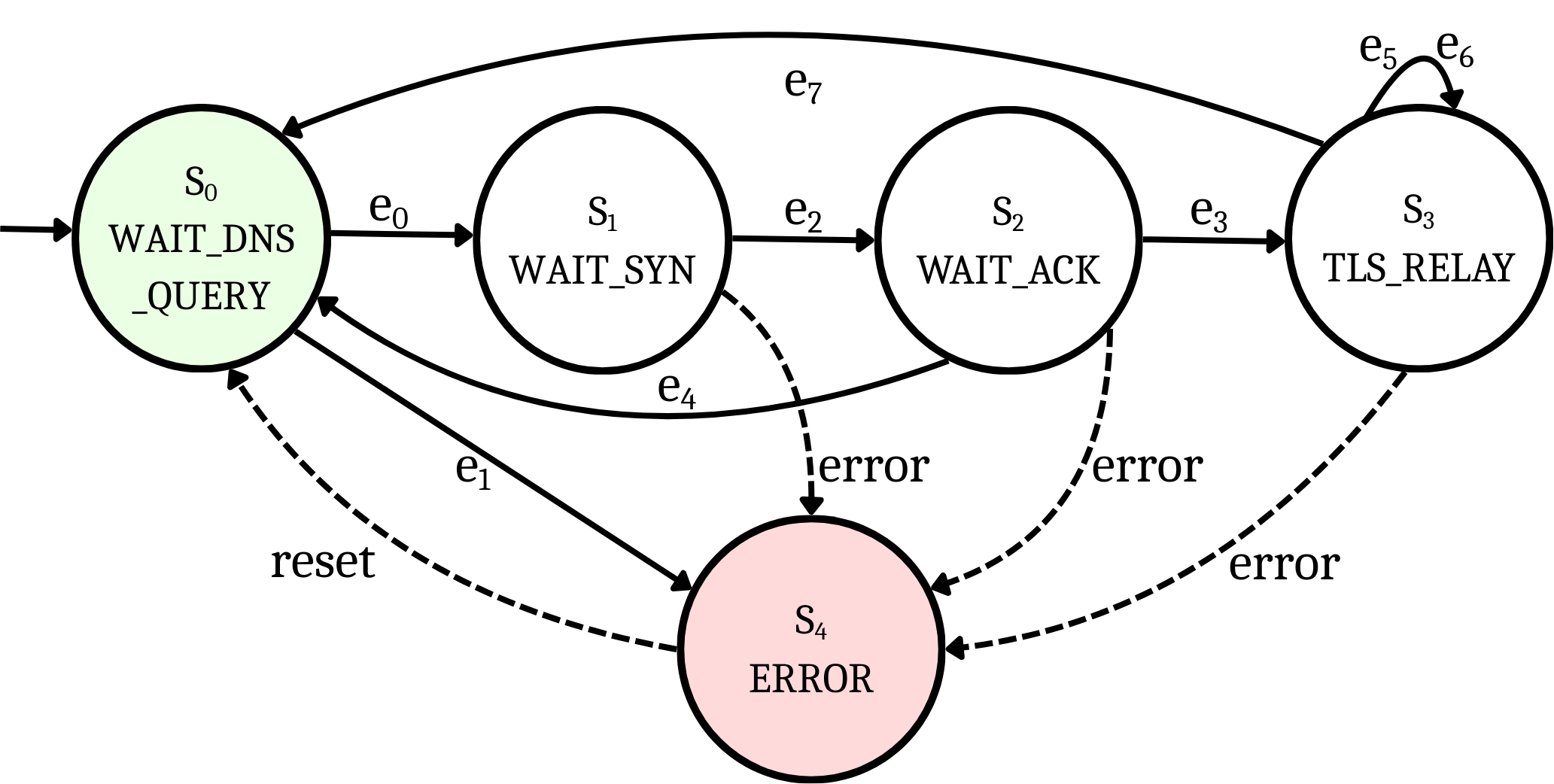}
    \caption{Net Relay State Machine}
    \label{fig:serverfsm}
\end{figure}

$e_0$: LoRa message with domain name received, $e_1$: DNS resolution failed, $e_2$: LoRa message with TCP SYN received, $e_3$: Received \emph{SYN-ACK} from web server, $e_4$: Timeout or failure receiving \emph{SYN-ACK}, $e_5$: LoRa message with final TCP ACK received, $e_6$: Received LoRa message with TLS fragment, $e_7$: Session end or failure. Figure \ref{fig:serverfsm} depicts the FSM and Table~\ref{tab:server_delta} depicts the state transitions.

\begin{table}[h!]
    \centering
    \caption{Transition Function ($\delta_s$) for the Server FSM.}
    \label{tab:server_delta}
    \scriptsize
    \begin{tabularx}{\columnwidth}{l l l X}
        \toprule
        \textbf{State} & \textbf{Event} & \textbf{Next State} & \textbf{Action(s)} \\
        \midrule
        $S_0$ & $e_0$ & $S_1$ & Resolve domain, send IP over LoRa \\
        $S_0$ & $e_1$ & $S_4$ & Report error and clean up \\
        $S_1$ & $e_2$ & $S_2$ & Modify and Send SYN to target \\
        $S_2$ & $e_3$ & $S_3$ & Correct TCP timestamp and send SYN-ACK over LoRa \\
        $S_2$ & $e_4$ & $S_0$ & Reset and report error \\
        $S_3$ & $e_5$ & $S_3$ & Forward ACK to target server \\
        $S_3$ & $e_6$ & $S_3$ & Send LoRa ACK, reassemble packet, and forward to target \\
        Any & $e_7$ & $S_0$ & Report error and clean up \\
        \bottomrule
    \end{tabularx}
\end{table}

\subsection{Threat Model, Assumptions, and Guarantees}
In the \emph{TLoRa} architecture, we consider a) Hotspot Adversary ($\mathcal{A}_{hotspot}$) who can potentially intrude on the hotspot created by the End Hub ($\Psi_c$) and may cause harm. b) LoRa Adversary ($\mathcal{A}_{lora}$) who can operate on the LoRa link between the $\Psi_c$ and the Net Relay ($\Psi_s$). This entity may modify packets, replay previous packets, and may jam the link.

We assume the Internet-based adversary to be a standard Dolev-Yao one. Its adverse attacks are mitigated by the already secure and correctly implemented TLS 1.3 protocol. Moreover, the $\Psi_c$, $\Psi_s$, and End Devices (($s_i \in S$)) are trusted and their hardware and software, including the \emph{TLoRa} code, are not compromised.

\emph{TLoRa} ensures end-to-end guarantees inherited from the TLS protocol tunnel. Since the TLS session is not terminated at the proxies ($\Psi_c$ and $\Psi_s$), the adversary $\mathcal{A}_{lora}$ can only see cyphertexts. An attempt to modify a payload would result in corruption of the TLS record. Furthermore, the identity of the participating entities is ensured by the certificates in the transparent tunnel.

\subsection{Admission Control Model}
\emph{TLoRa} implements a \emph{Sentinel} that prevents the constrained LoRa channel from oversubscription. It is conceptually a sub-state within the WAIT\_SYN state of the EH's FSM. As the FSM detects a TCP SYN at time $t$, it invokes the \emph{Sentinel} to enforce two admission control policies. The first policy is to check for concurrency given by -

\begin{equation}
    N_{active}(t) < N_{max}
\end{equation}

where, $N_{active}(t)$ is the number of currently active sessions and $N_{max}$ is the maximum number of allowed simultaneous sessions.

The second policy controls the rate at which new connections are introduced in the system using a token bucket to control the average rate of new connections. It protects the system from overloading due to a sudden increase in the number of new connections. This policy ensures that at least one token is available at a given time and the token generation method is represented as - 

\begin{equation}
T(t) = \min(T_{max}, T(t-\Delta t) + \rho \cdot \Delta t)
\end{equation}

where $T(t)$ is the number of tokens present in the bucket at a given time $t$, $T_{max}$ is the capacity of the bucket, $\rho$ is the rate at which new tokens are generated, and $\Delta t$ is the elapsed time since the last request. The \emph{Sentinel} allows a session only if both conditions are met. The final decision is given by -

\begin{equation}
    Admit(t) =
    \begin{cases}
    True & \text{if } N_{active}(t) < N_{\max} \land T(t) \geq 1, \\
    False & \text{otherwise}.
    \end{cases}
\end{equation}

\emph{TLoRa} drops the SYN packet if the corresponding session is rejected. The client may retry later. The \emph{Sentinel} handles concurrency elegantly in the \emph{TLoRa} system.

\section{Implementation}
We implemented and tested the proposed \emph{TLoRa} system in real-time in a laboratory environment. The EH ($\Psi_c$) consisted of a Raspberry Pi 3B+ single board computer (SBC) and an RFM95W LoRa transceiver. The SBC consisted of a Cortex-A53 (ARMv8) 64-bit SoC @ 1.4GH$_z$ and
1GB RAM. The LoRa transceiver operated at 866MH$_z$ with a 5dBi gain antenna. It was configured to use Spreading Factor (SF) 7 and Bandwidth (BW) 500KH$_z$. These configurations for LoRa was to ensure the keep the latency to the minimum possible. The NR ($\Psi_s$) also had a similar setup with the SBC being a Raspberry Pi 5 with a 2.4GHz quad-core 64-bit ARM Cortex-A76 CPU and 4GB RAM. Both the SBCs ran 64-bit Raspberry Pi OS (Bookworm) and Python v3.11.2. $\Psi_s$ connected to a home WiFi router with Internet connection. We tested \emph{TLoRa} with a mock API provided created on Beeceptor and our own Django web server hosted on an AWS EC2 instance with TLS 1.3 configured.

The Scapy Python library was used for the packet handler in our implementation, as the Threading and Queue libraries managed concurrent operations like sniffing and LoRa communication. Hostapd and dnsmasq Linux utilities were used to create and manage the WiFi hotspot and the captive portal. Additionally, we modified the pyLoraRFM9x library to suit our requirements in the implementation. 

The algorithm \ref{eh_session_start} shows the procedure when the EH captures a TCP SYN packet from an end device and proceeds to initiate a TCP handshake. The algorithm executes in $\mathcal{O}(1)$ time as it handles a single event and wait time dominates its lifetime.

\begin{algorithm}
\caption{EH Session Start and Handshake}
\label{eh_session_start}
\begin{algorithmic}[1]
    \renewcommand{\algorithmicrequire}{\textbf{Inputs:}}
    \renewcommand{\algorithmicensure}{\textbf{Output:}}
    \Require{$P_{syn} \gets$ TCP SYN packet, $Q_s$ (session queue)} 
    \Ensure{$G_{syn} \gets$ status (\texttt{Success} or \texttt{Failure})}

    \Procedure{Handshake}{$P_{syn}$, $Q_{s}$}
        \If{$P_{syn}$}
            \State{$Q_{s} \gets P_{syn}$}
            \State{$W_s \gets Q_{s}.\texttt{DEQUEUE}()$}
            \State{Send $W_s$ over LoRa}
            \State{$W_{S_{ack}}.\texttt{WAIT(timeout)}$}
            \If{$W_{S_{ack}}$}
               \State{$W_{ack} \gets \texttt{SEND\_CAPTURE\_ACK($W_{S_{ack}}$)}$}
            \EndIf
            \If{$W_{ack}$}
               \State{Send $W_{ack}$ over LoRa to NR}
               \State{\textbf{return} $G_{syn} \gets \texttt{Success}$}
            \EndIf            
        \EndIf
    \EndProcedure
\end{algorithmic}
\end{algorithm} 

The NR implements algorithm \ref{alg:nr_handshake} to act as a stateful proxy and reconstruct the TCP handshake. It performs cross-layer modification of the TCP timestamp in the server's SYN-ACK response. This is the most critical function to ensure a successful handshake in the \emph{TLoRa} system's high latency LoRa link. The algorithm executes in $\mathcal{O}(1)$ time as it processes a fixed-sized message.

\begin{algorithm}
\caption{NR TCP Handshake Reconstruction}
\label{alg:nr_handshake}
\begin{algorithmic}[1]
    \renewcommand{\algorithmicrequire}{\textbf{Inputs:}}
    \renewcommand{\algorithmicensure}{\textbf{Output:}}
    \Require{$M$ (LoRa Message), $S$ (FSM State)}
    \Ensure{$G_{recon} \gets$ status (\texttt{Success} or \texttt{Failure})}

    \Procedure{HandleHandshakeMessage}{$M, S$}
        \If{$S$ is \texttt{WAIT\_SYN}}
            \State $P_{syn} \gets \texttt{PARSE\_IP}(M)$
            \State $tsval_{orig} \gets \texttt{GETCLIENTTIMESTAMP}(P_{syn})$
            \State $P_{syn\_ack} \gets \texttt{SYNTOSERVER\_AWAITRESP}(P_{syn})$
            
            \If{$P_{syn\_ack}$}
                \State \textbf{OverwriteTimestamp($P_{syn\_ack}$, $tsval_{orig}$)}
                \State Send corrected $P_{syn\_ack}$ over LoRa
                \State $S \gets \texttt{WAIT\_ACK}$ 
                \State \textbf{return} $G_{recon} \gets$ \texttt{SUCCESS}
            \EndIf
        \ElsIf{$S$ is \texttt{WAIT\_ACK}}
            \State $P_{ack} \gets \texttt{PARSE\_IP}(M)$
            \State Forward $P_{ack}$ to Web Server
            \State $S \gets \texttt{TLS\_RELAY}$ \small {\Comment{Update state, handshake done}}
            \State \textbf{return} $G_{recon} \gets$ \texttt{SUCCESS}
        \EndIf
        \State \textbf{return} $G_{recon} \gets$ \texttt{FAILURE}
    \EndProcedure
\end{algorithmic}
\end{algorithm}

We evaluated the performance of the \emph{Sentinel} by simulating concurrent requests and recording the KPIs which were independent of the variability in the LoRa physical layer. In our experiments, we set the number of clients to be $20$ and varied the client arrival rate. The three rounds of the experiment captured the behavior of the system under low, medium, and high client arrival rates. Throughout the experiment, the maximum number of allowed client was set to $1$ and one token in the bucket was generated every $15$ seconds.

\section{Performance Evaluation}
\label{performance_evaluation}
We evaluate the implemented \emph{TLoRa} system by accessing an API twenty times over HTTPS. For each access, the performance metrics were recorded. The energy consumption by the EH and NR was also recorded using a USB energy meter. During the experiment, EH and NR were placed 10-12 meters apart as the study aimed to determine the baseline evaluation of the proposed system in a control environment.

\subsection{DNS Resolution Delay}
\begin{figure}[H]
    \centering
    \includegraphics[width=0.95\columnwidth]{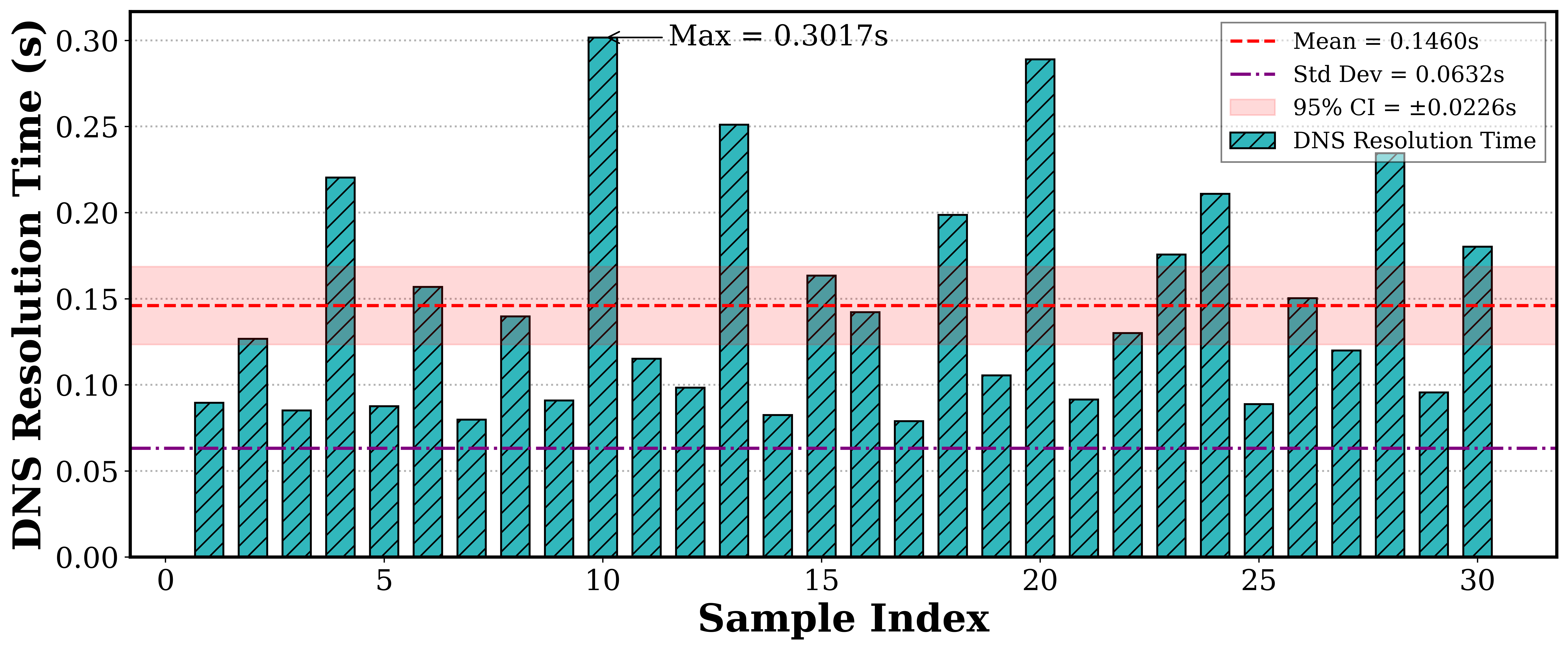}
    \caption{Time taken to resolve a domain name in \emph{TLoRa}}
    \label{fig:dns_resolution}
\end{figure}

We measured the DNS Resolution Time ($\Delta_{DNS}$) by requesting a URL $30$ times. Figure \ref{fig:dns_resolution} depicts the observations along with the maximum, the mean, and the standard deviation, which were $0.302$s, $0.146$s, and $0.063$s, respectively. Prior work reported that $85\%$ of UDP A-record lookups completed within $250$ms. However, only $42-49\%$ encrypted transports successfully completed within the $250$ms and required $41-44$s for $99\%$ completion \cite{10.1145/3609423}. The results are in stark comparison against our results, where the DNS lookup only took $0.146$s on average, and the total HTTPS request completion time was under only $14$s, as in Section \ref{total_delay}.

\subsection{TCP Handshake Delay}
\begin{figure}[H]
    \centering
    \includegraphics[width=0.95\columnwidth]{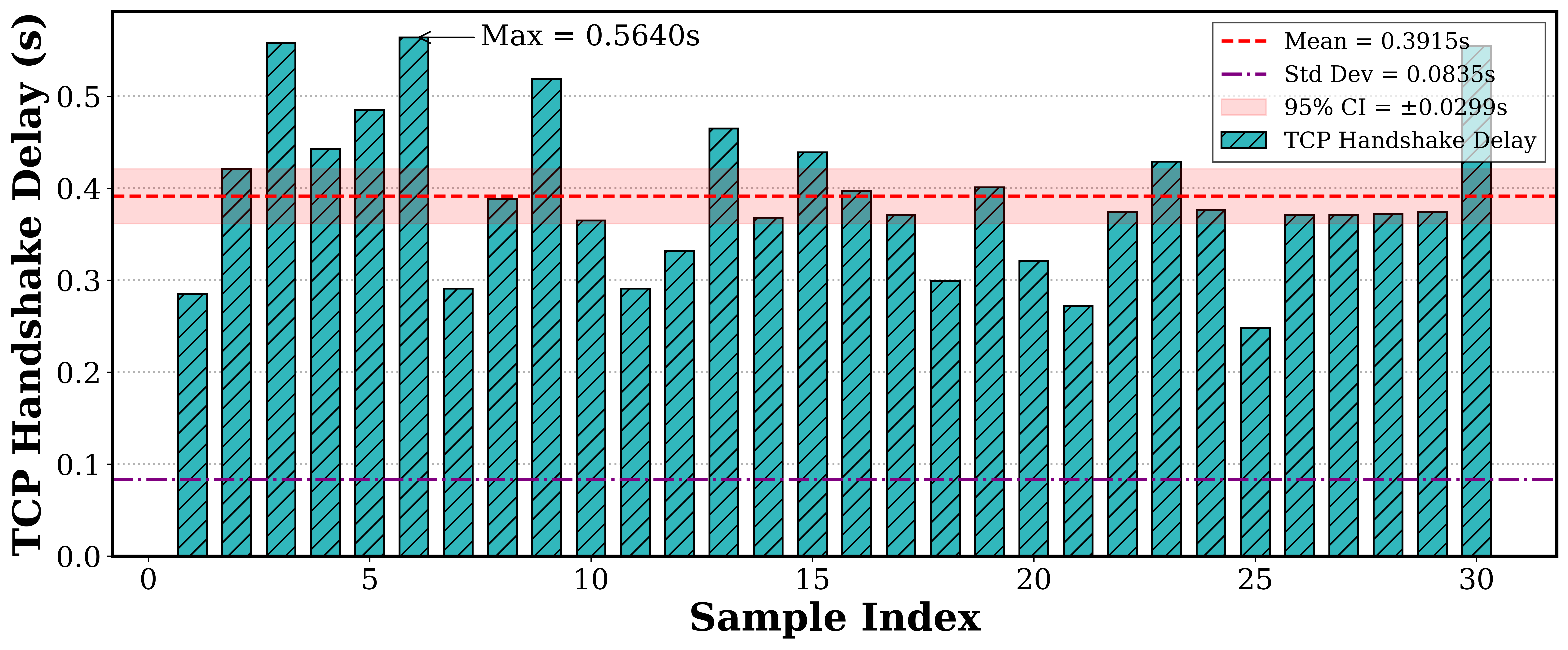}
    \caption{Time taken for three-way TCP handshake in \emph{TLoRa}}
    \label{fig:tcp_handshake}
\end{figure}

Figure \ref{fig:tcp_handshake} shows the results obtained from our experiments with the \emph{TLoRa} system. The average time for a complete three-way TCP handshake ($\Delta_{TCP}$) was observed to be $0.3915$s. A standard deviation of $0.0835$s was also observed in the $30$ URL requests made during the experiment.

\subsection{TLS Handshake Delay}
\begin{figure}[H]
    \centering
    \includegraphics[width=0.95\columnwidth]{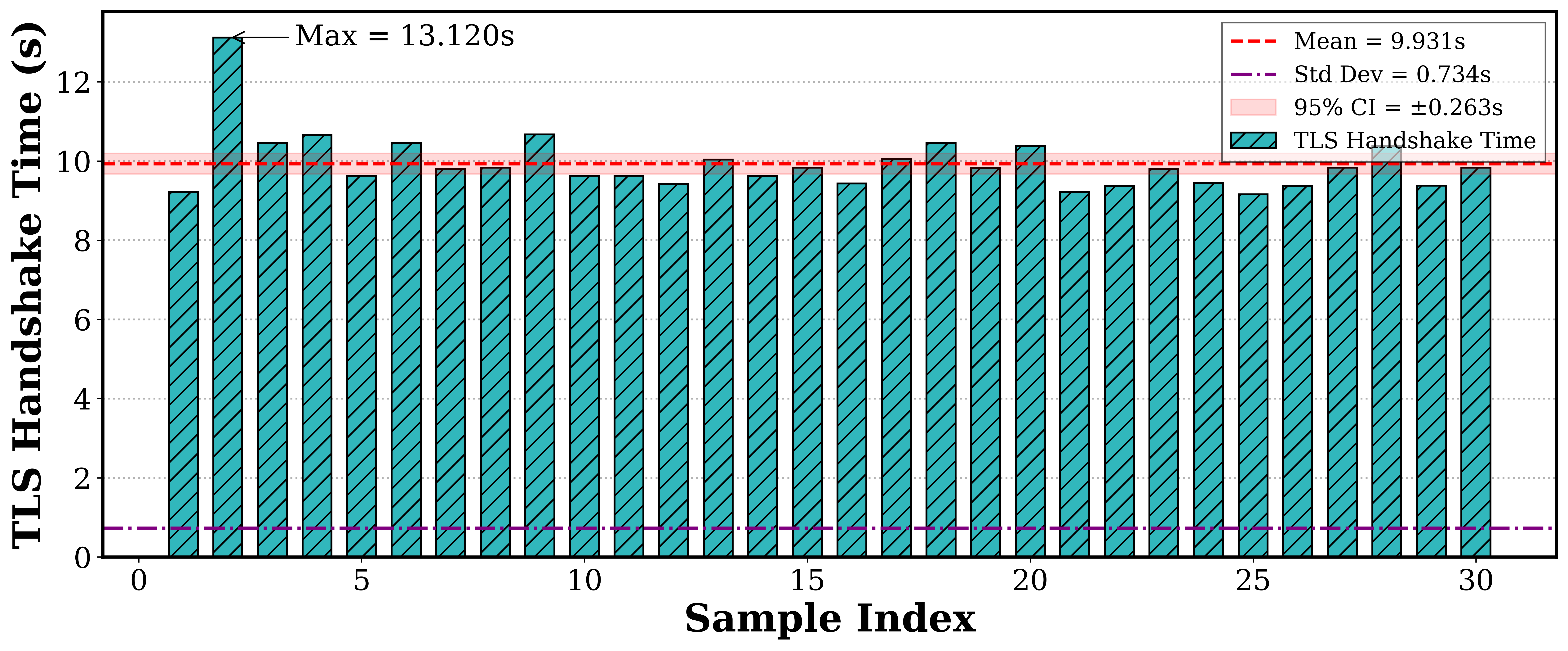}
    \caption{Time taken for establishing TLS connection in \emph{TLoRa}}
    \label{fig:tls_handshake}
\end{figure}

The TLS handshake time ($\Delta_{TLS}$) was also recorded to gauge the performance of the proposed \emph{TLoRa} system. \emph{TLoRa} outperformed the implementation of TLS over LoRaWAN in \cite{9861875}, which took approximately $12$s, whereas the TLS handshake in \emph{TLoRa} only took an average of $9.9$s. Furthermore, our work performed orders of magnitude better than the certificate-based DTLS handshake over LoRaWAN as reported in \cite{mathew2025realistic}. Interestingly, the TLS handshake time in \emph{TLoRa} implementation was very similar to the handshake time for the optimized PSK implementation. Figure \ref{fig:tls_handshake} depicts the observations from our experiment.

\subsection{API Access Delay}
\begin{figure}[H]
    \centering
    \includegraphics[width=0.95\columnwidth]{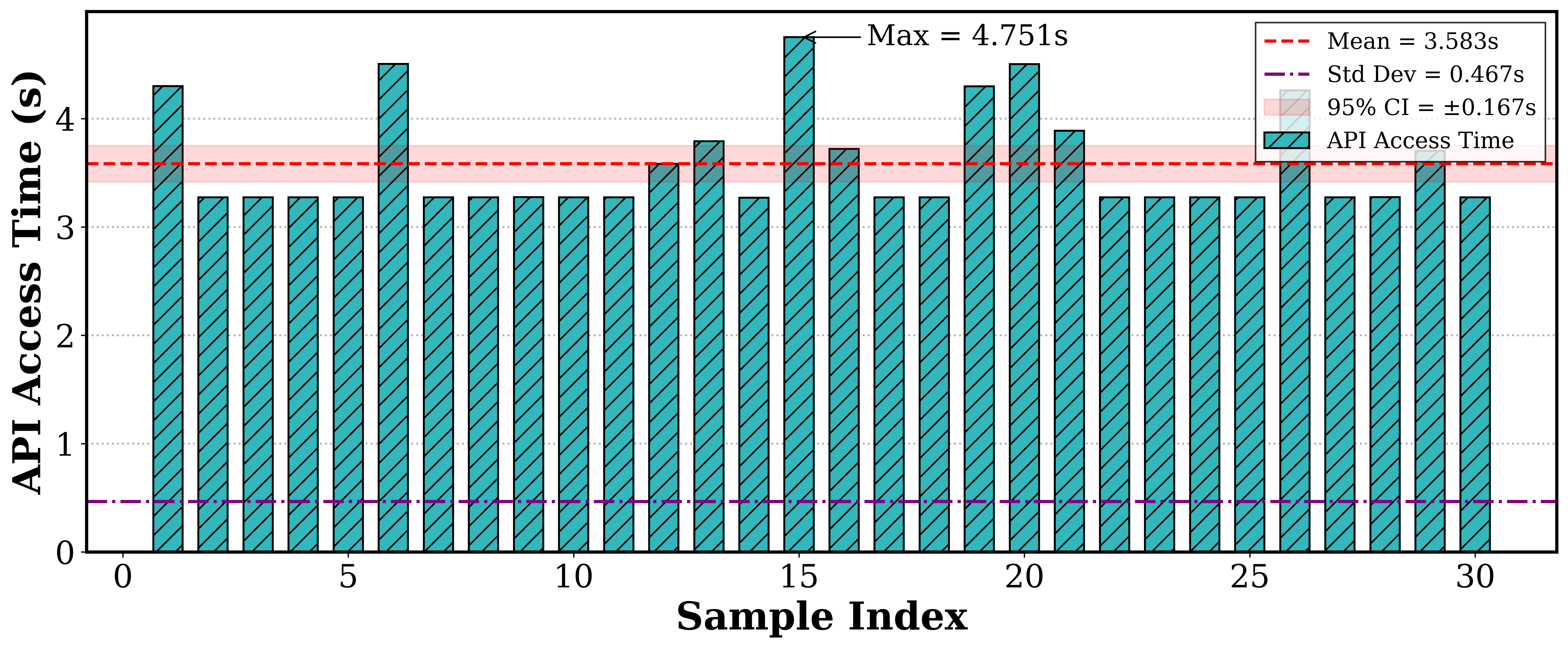}
    \caption{Time taken for completing a \texttt{GET} request}
    \label{fig:get_transfer}
\end{figure}

We recorded the API Access Delay ($\Delta_{Access}$), which is the total time it took for a \texttt{GET} request to be sent to the web server and receive the complete resource at the end device. The URL returned a $55$ byte-sized JSON content. On average, $\Delta_{Access}$ was observed to be $3.583\pm0.467$s. Figure \ref{fig:get_transfer} depicts the results from our experiment.

\subsection{Total Delay}
\label{total_delay}
The end-to-end request fulfillment in the \emph{TLoRa} system is a set of activities - $\mathcal{B} = \{\Delta_{DNS}, \Delta_{TCP}, \Delta_{TLS}, \Delta_{\texttt{Access}}\}$. The total delay is given by -

\begin{equation}
    \label{delta_total}
    \Delta_{\text{total}} = \sum_{i \in \mathcal{B}} \Delta_i
\end{equation}

\begin{figure}[H]
    \centering
    \includegraphics[width=0.95\columnwidth]{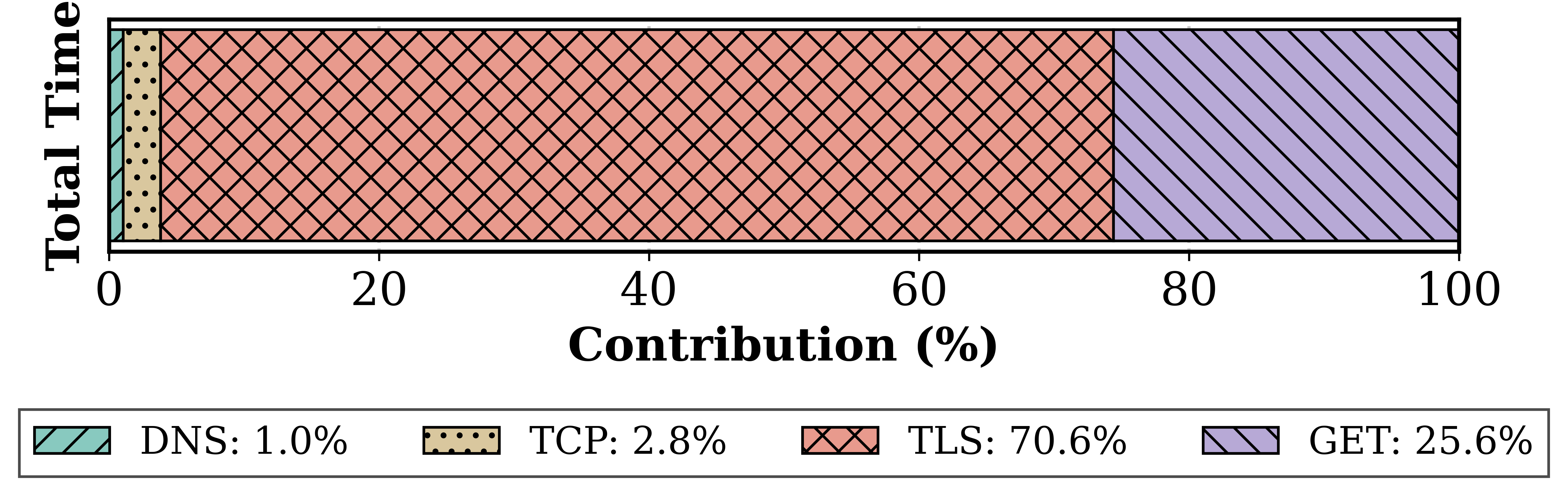}
    \caption{Time taken to complete an HTTPS request in the \emph{TLoRa} system}
    \label{fig:total_delay}
\end{figure}

In our experiments, we observed the average $\Delta_{total}$ to be $14.02 \pm 2.05$s. The $\Delta_{total}$ was dominated by the $\Delta_{TLS}$, which constituted $\approx71\%$ of the total time. Figure \ref{fig:total_delay} depicts the contribution of each stage of the HTTPS request. Our end-to-end delay for an HTTPS URL request is $\approx13.72$s when compared to the $\approx3-5$s airtime as calculated by Rademacher et al. \cite{9861875} for $SF = 7$ indicates that $\approx 5-8$ seconds of additional delay were introduced by device processing and network overheads, components not included in their model.

\subsection{Packet Delivery Ratio (PDR) and Throughput}
To estimate the quality of the \emph{TLoRa} network, we calculate the Packet Delivery Ratio (PDR ($\eta$)). It is the number of packets delivered ($\beta_{r}$) to the total number of packets sent ($\beta_s$). Since our implementation included an acknowledgment and retry mechanism, all the packets were delivered. Thus -

\begin{equation}
    \eta [\%] = \frac{\beta_r}{\beta_s} \times 100 = 100\%
\end{equation}

We measured the throughput of the system by transferring a $2KB$ file from the web server to the end device (a laptop) connected to the End Hub's WiFi hotspot. A custom Python script was used to send a \texttt{GET} request to the web server, and the transfer times were recorded. The average transfer time was observed to be $5.73$s. The average throughput was $357.42$ B/s.

\subsection{CPU, RAM, and Energy Consumption}
The KPIs, such as the RAM, CPU, and the energy consumption of both the EH and the NR, were recorded while accessing the HTTPS URL. The EH being deployed on a Raspberry Pi 3B+ device, which was a significantly lower-performing device than the Raspberry Pi 5 of the NR, consumed more CPU and RAM. It consumed $\approx83\%$ of CPU as compared to $\approx60\%$ in the NR. Additionally, it consumed about $4.5\%$ more RAM than the Raspberry Pi 5-based NR. But, the Raspberry Pi 5 consumed more energy due to its powerful CPU and onboard fan.

\begin{figure}[H]
    \centering
    \includegraphics[width=0.95\columnwidth]{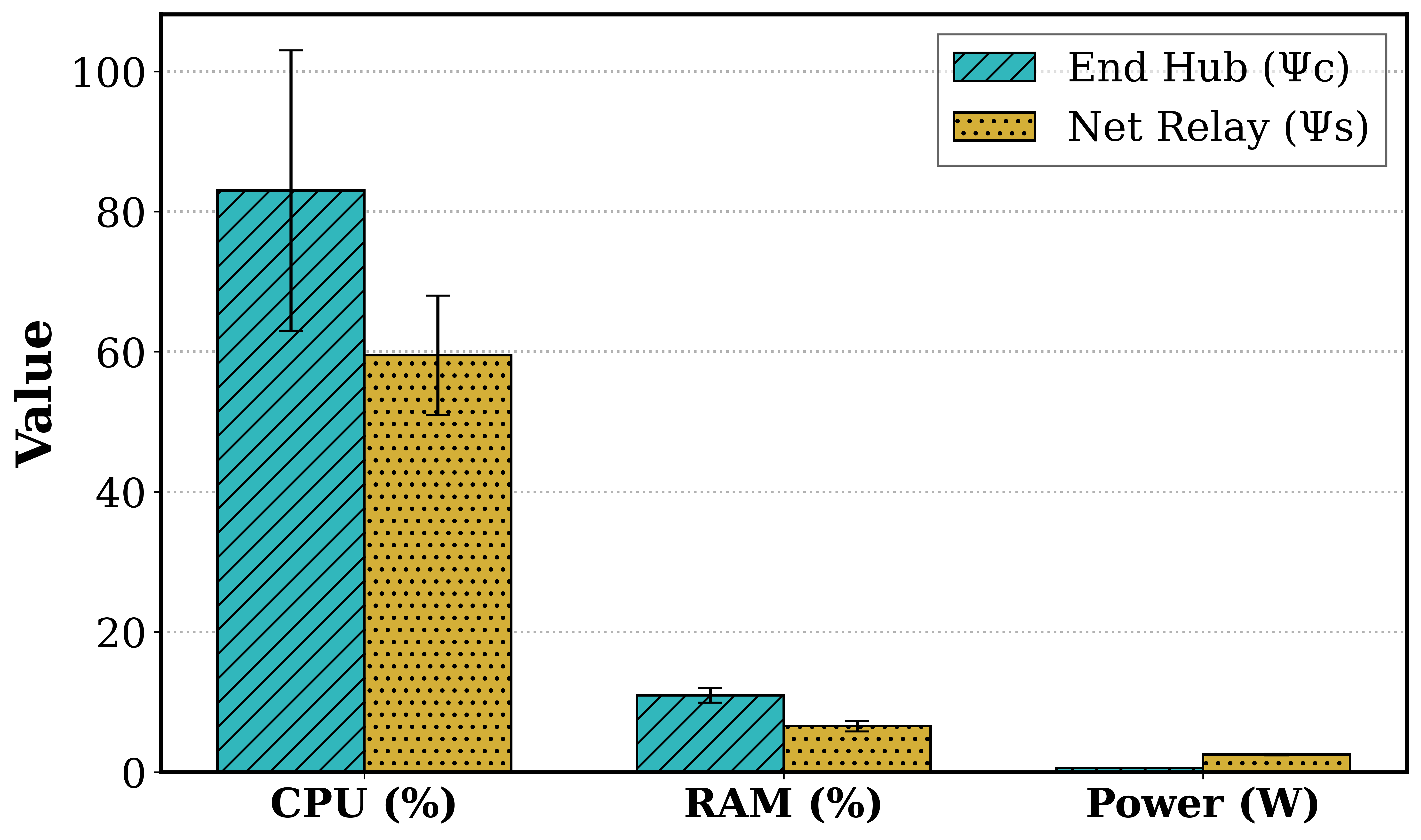}
    \caption{RAM, CPU, and Power consumption by End Hub and Net Relay}
    \label{fig:cpu_ram_power}
\end{figure}

\subsection{Duty Cycle Analysis}
With SF = $7$ and BW = $500KH_z$, we consider the \emph{TLoRa} system requests an API every $20$ minutes for the Duty Cycle (DC) calculations. Additionally, we consider that each request initiates a new TLS session. Hence, using $\Delta_{total}$ in Eq. \ref{delta_total} - 

\begin{equation*}
    \text{DC (\%)} = \frac{14}{1200} \times 100 \approx 1.17\%
\end{equation*}

\begin{equation}
     = 1.17\pm (\frac{2.05}{1200}\times100) \approx 1.17\pm0.17\%
\end{equation}

The frequency of the on-time can be easily adjusted to match applications and respect regional duty cycle requirements. However, with high spreading factors such as $9$ and $12$, the on time increases sharply and may require the user to adjust the frequency of the on events. 

\subsection{Sentinel Performance}
\begin{figure}[H]
    \centering
    \includegraphics[width=0.95\columnwidth]{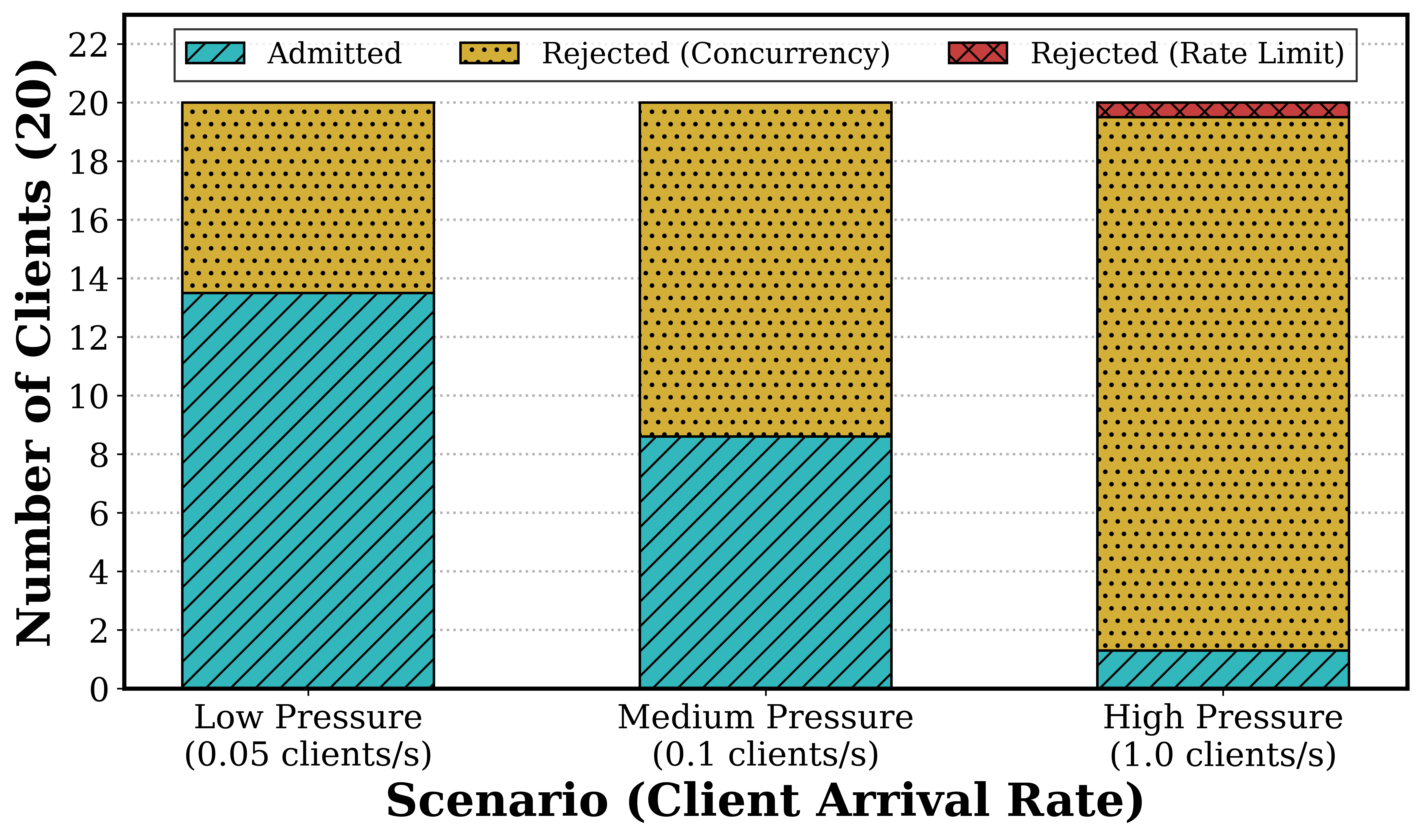}
    \caption{Summary of \emph{Sentinel} performance in our experiments}
    \label{fig:sentinel_graph}
\end{figure}

Table \ref{tab:table_sentinel} shows the \emph{Sentinel's} efficiency in handling concurrent requests and how it protects the LoRa channel from oversubscription. Figure \ref{fig:sentinel_graph} depicts the scenarios where the \emph{Sentinel} successfully throttled connection requests as the arrival rate increased. The maximum number of allowed connections (concurrency) was the main dominant reason for rejecting connections. The token-bucket's rate limit became a minor reason for flooding connections.

\begin{table}[h]
    \scriptsize
    \centering
    \caption{Summary of \emph{Sentinel} performance.}
    \label{tab:table_sentinel}
    \begin{tabularx}{\columnwidth}{@{}c c c c c@{}}
        \toprule
        \textbf{Scenario} & 
        \makecell{\textbf{Arrival} \\ \textbf{Rate} \\ \textbf{(clients/s)}} & 
        \makecell{\textbf{Average} \\ \textbf{Admitted} \\ ($\pm$ Stdev)} & 
        \makecell{\textbf{Average} \\ \textbf{Rejected} \\ \textbf{(Concurrency)}} & 
        \makecell{\textbf{Average} \\ \textbf{Rejected} \\ \textbf{(Rate Limit)}} \\
        \midrule
        Low     & 00.05 & 13.50 $\pm$ 01.20 & 06.50 $\pm$ 01.20 & 00.00 $\pm$ 00.00 \\
        Medium  & 00.10 & 08.60  $\pm$ 00.50 & 11.40 $\pm$ 00.50 & 00.00 $\pm$ 00.00 \\
        High    & 01.00 & 01.30 $\pm$ 00.50 & 18.20 $\pm$ 00.80 & 00.50 $\pm$ 00.50 \\
        \bottomrule
    \end{tabularx}
\end{table}

\section{Conclusion and Limitations}
In this work, we proposed \emph{TLoRa}, a flexible and extensible architecture to enable secure and direct API access over LoRa using TLS. The end devices and the user's application do not require any modification to adopt \emph{TLoRa}. We also evaluated TLoRa by implementing a lab-scale prototype on actual hardware.

The proposed \emph{TLoRa} system currently only supports API access. Multimedia web page access on the TLoRa system would require advanced methods, which we plan to take up as future work.  Furthermore, we also plan to evaluate the scalability of the TLoRa system over a large-scale deployment.

\bibliographystyle{IEEEtran}
\bibliography{IEEEabrv, references}

\end{document}